\documentclass[12pt]{article}
\usepackage{cite}

\def\X#1{\mbox{\raisebox{1.2pt}{$\stackrel{#1}{X}$}}}

\begin{document}

\begin{center}
{\Large \bf Method of group foliation, hodograph transformation
and non-invariant solutions \protect\\[2mm] of the Boyer-Finley equation}\\[2mm]
{\large \bf M B Sheftel$^{1,2}$}  \\[1mm]
$^1$ Feza G\"{u}rsey Institute, PO Box 6, Cengelkoy, 81220
Istanbul, Turkey
\\ $^2$ Department of Higher Mathematics, North Western State
Technical University, Millionnaya St. 5, 191186, St. Petersburg,
Russia
\\ E-mail: sheftel@gursey.gov.tr
\end{center}

\date{}

\vspace{1mm}

\noindent  {\bf Abstract} \\

We present the method of group foliation for constructing
non-invariant solutions of partial differential equations on an
important example of the Boyer-Finley equation from the theory of
gravitational instantons. We show that the commutativity
constraint for a pair of invariant differential operators leads to
a set of non-invariant solutions of this equation. In the second
part of the paper we demonstrate how the hodograph transformation
of the ultra-hyperbolic version of Boyer-Finley equation in an
obvious way leads to its non-invariant solution obtained recently
by Ma\~nas and Alonso. Due to extra symmetries, this solution is
conditionally invariant, unlike non-invariant solutions obtained
previously. We make the hodograph transformation of the group
foliation structure and derive three invariant relations valid for
the hodograph solution, additional to resolving equations, in an
attempt to obtain the orbit of this solution.

\vspace{2mm}



\section{Introduction}
\label{sec:intro}

 The standard method for obtaining exact solutions of partial
differential equations (PDEs) by the symmetry group analysis is
the symmetry reduction which gives only {\it invariant solutions},
{\it i.e.} the solutions which are invariant with respect to some
subgroup of symmetry group of the PDEs \cite{ovs,olv,wint}. The
drawback of the standard symmetry analysis is that with this
method one loses all non-invariant and many invariant solutions,
the latter being non-invariant solutions of the equations obtained
by symmetry reduction.

 Meanwhile, non-invariant solutions of PDEs might be very important for
applications in physics. The classical example is the so-called
$K3$ instanton in the Einstein's theory of gravitation which
corresponds to the Kummer surface in differential geometry. The
corresponding K\"ahler metric has no Killing vectors and it could
be constructed once one finds non-invariant solutions of the
complex Monge-Amp\`ere equation, the problem which has not been
solved for over $150$ years.

 Recently we discovered that the method of group foliation
can be made an appropriate tool for obtaining non-invariant
solutions of non-linear PDEs that admit an infinite-dimensional
symmetry group. The method involves the study of compatibility of
given equations with a differential constraint, which is
automorphic under an infinite dimensional symmetry subgroup, the
latter acting transitively on the submanifold of the common
solutions. By studying compatibility conditions of this
automorphic system, {\it i.e.} the resolving equations, one can
provide an explicit foliation of the entire solution manifold of
the considered equations into separate orbits. Unlike the standard
symmetry method we do not loose any solutions in the process and
we can select orbits of non-invariant solutions by obtaining
appropriate solutions of the resolving system.

 The idea of the method, belonging to Lie and Vessiot \cite{lie,vessiot},
is more than a hundred years old being resurrected in a modern
form by Ovsiannikov more than $30$ years ago (see \cite{ovs} and
references therein). We have modified the method by introducing
three important new ideas \cite{ns,nstmp,msw,sh1,sh2}:
\\ 1. The use of {\it invariant cross-differentiation,} involving
operators of invariant differentiation and their commutator
algebra for derivation of resolving equations and for obtaining
their particular solutions.
\\ 2. {\it Commutator representation} of resolving system in terms
of the commutators of operators of invariant differentiation.
\\ 3. Method of {\it invariant integration} for solving automorphic
system.

We illustrate all this on a physically important example of the
Boyer-Finley equation \cite{BoyFin} which is known also as the
heavenly equation
\begin{equation}
u_{z\bar z}=\kappa (e^u)_{tt} \quad\iff\quad u_{xx}+u_{yy}=\kappa
(e^u)_{tt}, \qquad \kappa=\pm 1 \label{heav}
\end{equation}
where $u=u(t,z,\bar z)$. This equation is a continuous version of
the Toda lattice or $SU(\infty)$ Toda field \cite{stra}. It
appears in the theory of gravitational instantons
\cite{eguchi,tod,nush} where it describes self-dual Einstein
spaces with Euclidean signature having one rotational Killing
vector. It also appears in the general theory of relativity
\cite{ward} and other physical theories.

  In the first part of this paper, on the example of the Boyer-Finley
equation, we clarify main concepts of the method including these
three ideas and consider in detail the main steps which should be
performed for obtaining non-invariant solutions. This approach
enables us to present explicitly a family of non-invariant
solutions of the Boyer-Finley equation depending on two functional
parameters.

In the second part of the paper we discuss a new non-invariant
solution of the ultra-hyperbolic version of the Boyer-Finley
equation (\ref{heav}) with $\kappa=1$ obtained recently by Ma\~nas
and Alonso by the hodograph transformation of associated systems
of the hydrodynamic type \cite{manal} (see also \cite{fer} for a
generalization of the approach of \cite{manal}). We point out that
if one makes the hodograph transformation of the Boyer-Finley
equation itself, then it becomes obvious how to impose a
differential constraint immediately leading to the solution of
Ma\~nas and Alonso without any use of hydrodynamical systems or
the general theorem presented in \cite{manal}. We show that this
solution possesses extra symmetries depending on eight arbitrary
functions of $u$ which are not the symmetries of the Boyer-Finley
equation. Hence, this solution is conditionally invariant
\cite{levwin} though not an invariant solution. A similar argument
shows that symmetries of the solutions obtained by Calderbank and
Tod \cite{tod} and by Martina, Sheftel and Winternitz \cite{msw}
coincide with the symmetries of the Boyer-Finley equation itself
and hence these solutions, having no extra symmetries, are not
conditionally invariant. Finally, we make the hodograph
transformation of the whole group foliation structure and obtain
the relations between differential invariants valid for the
solution of \cite{manal} which together with the resolving
equations partially describe the orbit of this solution. We have
not yet succeeded to obtain a solution of the resolving system
which corresponds to the orbit of the hodograph solution of
\cite{manal}. Since the additional relations which we have found
do not form a complete set of invariant differential constraints
which uniquely select the hodograph solution, the corresponding
solution of the resolving system would give us the orbits of more
general solutions of (\ref{heav}) with $\kappa=1$ and also
solutions with $\kappa=-1$. This work is still in progress.

\section{Symmetry algebra, differential invariants and automorphic system}
\setcounter{equation}{0}

 We start with the {\it symmetry algebra} of the generators of
point transformations for the Boyer-Finley equation (\ref{heav})
\cite{msw}
\begin{eqnarray}
T = \partial_t, \qquad G=t\partial_t+2\partial_u \nonumber\\ X_a=
a(z)\partial_z +\bar a(\bar z)\partial_{\bar z} -(a^\prime(z)+\bar
a^\prime(\bar z))\partial_u \label{symgen}
\end{eqnarray}
where $a(z)$ is an arbitrary holomorphic functions of $z$ and
prime denotes derivative with respect to argument.

 The Lie algebra of symmetry generators (\ref{symgen}) is determined by the
commutation relations
\begin{equation}\label{algeb}
[T,G]=T,\quad [T,X_a]=0,\quad [G,X_a]=0, \quad
[X_a,X_b]=X_{ab^\prime-ba^\prime}
\end{equation}
They show that the generators $X_a$ of conformal transformations
form a subalgebra of Lie algebra (\ref{algeb}). This subalgebra is
infinite dimensional since the generators $X_a$ depend on $a(z)$.
The corresponding finite transformations form an infinite
dimensional symmetry subgroup of the equation (\ref{heav}) since
instead of a group parameter they also involve an arbitrary
holomorphic function of $z$. We choose this infinite dimensional
{\it conformal group} for the group foliation.

 Next we find {\it differential invariants} of the symmetry subgroup
of conformal transformations. They are invariants of all the
generators $X_a$ in the {\it prolongation spaces}, so that they
can depend on independent variables, unknowns and also on partial
derivatives of unknowns allowed by the prolongation order. The
{\it order} of the differential invariant is defined as the order
of the highest derivative which this invariant depends on. The
highest order $N$ of differential invariants should be larger or
equal to the order of the equation $(N\ge 2)$ and there should be
$n$ functionally independent invariants with $n > p+q$ where $p$
and $q$ are the number of independent and dependent variables,
respectively. Thus, we should have $p=3,\; q=1$, $n > 4$, $N\ge
2$.

The routine calculation for $N=2$ gives $n=5$ functionally
independent differential invariants up to the second order
inclusively
\begin{equation}\label{difinv}
t,\qquad u_t,\qquad u_{tt},\qquad \rho=e^{-u}u_{z\bar z},\qquad
\eta=e^{-u}u_{zt}u_{\bar zt}
\end{equation}
and all of them are real. This allows us to express the
Boyer-Finley equation (\ref{heav}) solely in terms of the
differential invariants $u_{tt}=\kappa\rho-u_t^2$.

   The next step is choose the general form of the automorphic system.
 We choose the invariants $t,u_t,\rho$ as new {\it invariant
independent variables,} the same number $p=3$ as in the original
equation (\ref{heav}), and require that the {\it remaining
invariants should be functions of the chosen ones}. This provides
us with the general form of the {\it automorphic system} that also
contains the studied equation expressed in terms of invariants
(\ref{difinv})
\begin{equation}\label{autom}
u_{tt}=\kappa\rho-u_t^2,\qquad \eta=F(t,u_t,\rho)
\end{equation}
The real function $F$ in the right-hand side should be determined
from the {\it resolving equations} which are compatibility
conditions of the system (\ref{autom}). Then the system
(\ref{autom}) will have {\it automorphic property}, {\it i.e.} any
of its solutions can be obtained from any other solution by an
appropriate transformation of the conformal group.

\section{Operators of invariant differentiation and
the basis of differential invariants}
\setcounter{equation}{0}

Our next task is to find {\it operators of invariant
differentiation}.  They are linear combinations of total
derivatives $D_t$, $D_z$, $D_{\bar z}$ with respect to independent
variables $t,z,\bar z$ with the coefficients depending on local
coordinates of the prolongation space which are defined by the
special property that, acting on any (differential) invariant,
they map it again into a differential invariant. As a consequence,
these differential operators commute with any infinitely prolonged
generator $\X\infty_a$ of the conformal symmetry group. Being
first order differential operators, they raise the order of a
differential invariant by one.
 The total number of independent operators of invariant
differentiation is obviously equal to $3$, the number of
independent variables.

   We obtain the {\it basis for the operators of invariant differentiation}
\cite{msw,sh1,sh2}
\begin{equation}\label{deltas}
\delta=D_t,\qquad \Delta=e^{-u}u_{\bar zt}D_z,\qquad \bar
\Delta=e^{-u}u_{zt}D_{\bar z}
\end{equation}

The next step is to find the {\it basis of differential
invariants}\/ which is defined as a minimal finite set of
(differential) invariants of a symmetry group from which any other
differential invariant of this group can be obtained by a finite
number of invariant differentiations and operations of taking
composite functions. The proof of existence and finiteness of the
basis was given by Tresse \cite{tresse} and in a modern form by
Ovsiannikov \cite{ovs}.

  In our example the basis of differential invariants is formed by
the set of three invariants $t,u_t,\rho$, while two other
invariants $u_{tt}$ and $\eta$ defined in (\ref{difinv}) are given
by the relations
\begin{equation}\label{invdepend}
u_{tt}=\delta(u_t),\qquad \eta\equiv e^{-u}u_{zt}u_{\bar
zt}=\Delta(u_t) =\bar\Delta(u_t)
\end{equation}
All other functionally independent invariants of higher order can
be obtained by acting with operators of invariant differentiation
on the {\it basis} $\{t,u_t,\rho\}$.  In particular, the following
third order invariants generated from the second order invariant
$\rho$ by invariant differentiations will be used in our
construction
\begin{equation}\label{sigtau}
\sigma=\Delta(\rho),\quad \bar\sigma=\bar\Delta(\rho),\quad
\tau=\delta(\rho) \equiv D_t(\rho)
\end{equation}

\section{Commutator algebra of invariant
\protect\\ differential operators}
\setcounter{equation}{0}

 The operators $\delta$, $\Delta$ and $\bar\Delta$ defined by
(\ref{deltas}) form the {\it commutator algebra} which is a Lie
algebra over the field of invariants of the conformal group
\cite{ovs}.

  This algebra is simplified by introducing two new operators of
invariant differentiation $Y$ and $\bar Y$ instead of $\Delta$ and
$\bar\Delta$ and two new variables $\lambda$ and $\bar\lambda$
instead of $\sigma$ and $\bar\sigma$, defined by
\begin{equation}\label{Ydef}
\Delta=\eta Y,\quad \bar\Delta=\eta\bar Y,\quad
\sigma=\eta\lambda,\quad \bar\sigma=\eta\bar\lambda
\end{equation}
The resulting algebra has the form
\begin{eqnarray}
& & \bigl[\delta,Y\bigr]= \left(\kappa\bar\lambda
-3u_t-\frac{\delta(\eta)}{\eta}\right) Y, \qquad \bigl[\delta,\bar
Y\bigr]=\left(\kappa\lambda-3u_t-\frac{\delta(\eta)}{\eta}\right)\bar
Y \nonumber
\\ & &\bigl[Y,\bar Y\bigr]=
\frac{(\tau+u_t\rho)}{\eta} \left(Y - \bar Y\right)
\label{Yalg}
\end{eqnarray}

  With the use of operators $\delta$, $Y$ and $\bar Y$ the
general form (\ref{autom}) of the automorphic system becomes
\begin{equation}\label{modaut}
\delta(u_t) = \kappa\rho-u_t^2,\quad
Y(u_t) = 1 \qquad (\bar Y(u_t)=1)
\end{equation}
where the first equation is the Boyer-Finley equation and the
second equation follows from the second relation
(\ref{invdepend}). Here we put $\eta=F$ in the equations
(\ref{Ydef}) and in the commutation relations (\ref{Yalg})
according to the second equation in (\ref{autom}). Then we obtain
\begin{equation}\label{Yact}
Y=(1/F)\Delta,\quad \bar Y=(1/F)\bar\Delta,\quad
Y(\rho)=\lambda,\quad \bar Y(\rho)=\bar\lambda
\end{equation}

\section{Resolving equations}
\setcounter{equation}{0}

The system of {\it resolving equations}\/ is a set of
compatibility conditions between the studied equation and
equations added to it in order to form the automorphic system. In
our case we require the compatibility between two equations
(\ref{modaut}) which gives restrictions on the function
$F(t,u_t,\rho)$ in the right-hand side of the second equation in
(\ref{autom}). A new feature in our modification of the method is
that we do this in an explicitly invariant manner by using the
{\it invariant cross-differentiation} \cite{ns,nstmp,msw,sh1,sh2}
involving the operators $\delta$, $Y$ and $\bar Y$. The resulting
five resolving equations have the form \cite{msw,sh1,sh2}
\begin{eqnarray}\label{I}
\delta(F)=\Bigl[\kappa\Bigl(\lambda+\bar\lambda\Bigr)-5u_t\Bigr]F
\\[1mm] \label{II}
F\Bigl(Y(\bar\lambda)-\bar
Y(\lambda)\Bigr)=(\tau+u_t\rho)\bigl(\lambda-\bar\lambda\bigr)
\\[1mm] \delta(\lambda)=Y(\tau)+2u_t\lambda-\kappa\lambda^2
\label{III}\\[1mm]
\delta(\bar\lambda)=\bar
Y(\tau)+2u_t\bar\lambda-\kappa{\bar\lambda}^2
\label{bIII}
\\[1mm] \label{IV}
 F\Bigl(Y(\bar\lambda)+\bar Y(\lambda)\Bigr)=
-(\tau+u_t\rho)\bigl(\lambda+\bar\lambda\bigr) \nonumber
\\ \mbox{}+2\kappa\Bigl[\delta(\tau)+4u_t\tau+2F+\kappa\rho^2+2u_t^2\rho
\Bigr]
\end{eqnarray}
The definitions of $\lambda,\bar\lambda,\tau$ used here are given
by equations (\ref{Yact}), (\ref{sigtau}).

The resolving equations (\ref{I}), (\ref{II}),  (\ref{III}),
(\ref{bIII}), (\ref{IV}) form a closed {\it resolving system}
where the second order differential invariant $\eta=F$ and the
third order differential invariants $\lambda$, $\bar\lambda$ and
$\tau$ are unknown functions of the independent variables
$t,u_t,\rho$.

An explicit form of the equations (\ref{I}) --- (\ref{IV}) is
obtained if we use the {\it projected operators} of invariant
differentiation (see (\ref{deltas}), (\ref{modaut}), (\ref{Yact}))
\begin{equation}\label{deltproj}
\delta=\partial_t+(\kappa\rho-u_t^2)\partial_{u_t}+\tau\partial_\rho\,,
\quad Y=\partial_{u_t}+\lambda\partial_\rho\,,\quad \bar
Y=\partial_{u_t}+\bar\lambda\partial_\rho
\end{equation}

  The commutator relations (\ref{Yalg}) were satisfied identically
by the operators of invariant differentiation. On the contrary,
for the projected operators (\ref{deltproj}) these commutation
relations and even the Jacobi identity are satisfied only on
account of the resolving equations. More than that, the theorem
proved in \cite{msw} claims that commutator algebra (\ref{Yalg})
of the operators of invariant differentiation $\delta,Y,\bar Y$
together with the Jacobi identity gives a commutator
representation for the resolving system.

\section{Solutions of the resolving system
\protect\\ and of the Boyer-Finley equation}
\setcounter{equation}{0}

To find particular solutions of the resolving system, we impose
various simplifying constraints. The most obvious ones, like $\bar
Y=Y$ or $F=0$, lead to invariant solutions. These we already know,
or can obtain by much simpler standard methods. The weaker
assumption that leads to non-invariant solutions is that the
operators $Y$ and $\bar Y$ commute
\begin{equation}\label{Ycom}
[Y,\bar Y]=0 \quad\iff\quad \tau=-u_t\rho
\end{equation}
but $\bar Y\ne Y$, {\it i.e.} $\bar\lambda\ne\lambda$ and also
$F\ne 0$. With this Ansatz we find the {\it particular solution of
the resolving system} \cite{msw} (assuming $2\kappa\rho-u_t^2\ge
0$)
\begin{equation}\label{solresol}
\tau = -u_t\rho, \quad \lambda =\kappa u_t + i\sqrt{2\kappa\rho
-u_t^2}, \quad \bar\lambda =\kappa u_t - i\sqrt{2\kappa\rho
-u_t^2}
\end{equation}
and the expression for $F$ \cite{msw} is not needed for obtaining
solutions of (\ref{heav}).

To reconstruct solutions of the Boyer-Finley equation starting
from the particular solution (\ref{solresol}) of the resolving
system we use the procedure of {\it invariant integration} which
amounts to the transformation of equations to the form of {\it
exact invariant derivative} \cite{msw}. Then we drop the operator
of invariant differentiation in such an equation adding the term
that is an {\it arbitrary element of the kernel} of this operator.
This term plays the role of the integration constant. This
procedure is described in detail in \cite{msw,sh1,sh2}.

Here we present only the {\it final result} for solutions of the
Boyer-Finley equation \cite{msw} (see also \cite{tod} in relation
to (\ref{solminus}))
\begin{equation}\label{solplus}
u(t,z,\bar z) = \ln{\left|\frac{\bigl(t+b(z)\bigr)
c^\prime(z)}{c(z) + \bar c(\bar z)}\right|^2}\qquad {\rm for}\quad
\kappa=1
\end{equation}
\begin{equation}\label{solminus}
u(t,z,\bar z) = \ln{\left|\frac{\bigl(t+b(z)\bigr) c^\prime(z)}{1
+ |c(z)|^2}\right|^2}\qquad {\rm for}\quad \kappa=-1
\end{equation}
Here $b(z)$ and $c(z)$ are arbitrary holomorphic functions. One of
them is fundamental and labels a particular {\it orbit of
solutions}. The other one is induced by a conformal symmetry
transformation and can be transformed away.
 These solutions
for generic $b(z)$ and $c(z)$ are {\it non-invariant} \cite{msw}.

\section{Hodograph transformation of the Boyer-Finley
equation as a shortcut to its special solution}
\setcounter{equation}{0}

Hodograph transformation is the interchange of roles of the
unknown $u$ and one of the independent variables $t$:
$u=u(t,z,\bar z)\mapsto t=t(u,z,\bar z)$. The transformed equation
(\ref{heav}) has the form
\begin{equation}
\label{hodheav}
(t_zt_{\bar z}-\kappa e^u)t_{uu}=t_u(t_{\bar z}t_{uz}+t_zt_{u\bar z}
-t_ut_{z\bar z}-\kappa e^u)
\end{equation}
 There is
an obvious simplifying differential constraint
\begin{equation}
\label{ansatz} t_zt_{\bar z}-\kappa e^u = 0
\end{equation}
valid only for $\kappa=1$, since $t_zt_{\bar z}>0$. Indeed,
together with (\ref{hodheav}) it implies linear dependence of $t$
on $z$ and $\bar z$ and hence $t_z=e^{u/2+i\alpha(u)}$, $t_{\bar
z}=e^{u/2-i\alpha(u)}$ with the final result for the solution  of
(\ref{hodheav})
\begin{equation}
\label{spainsol}
t = e^{u/2+i\alpha(u)} z + e^{u/2-i\alpha(u)} \bar z + h(u)
\end{equation} \
where $\alpha(u)$, $h(u)$ are arbitrary functions.
 This is the solution given in \cite{manal}.

\section{Symmetries of non-invariant solutions}
\setcounter{equation}{0}

Point symmetries of the equation (\ref{hodheav}) coincide with the
symmetries (\ref{symgen}) of the original equation (\ref{heav})
because the hodograph transformation belongs also to a class of
point transformations. Since the solution (\ref{spainsol}) is the
general solution of the system of differential equations
(\ref{hodheav}), (\ref{ansatz}), the symmetries of this solution
coincide with the point symmetries of that system. A computer
package `Liepde' used together with the PDE solver `Crack' by
Thomas Wolf, that runs under REDUCE 3.6 or 3.7, gives the symmetry
generator of the solution (\ref{spainsol}) depending on eight
arbitrary functions of $u$
\begin{eqnarray}
\!\!&\! \!&\! X = a(u)\partial_u + \bigl[2zg(u)+2\bar
ze^ub(u)+te(u)+h(u)\bigr]\partial_t
\label{symsol}
\\ \!\!&\! \!&\! \mbox{} + \bigl[tb(u)+zc(u)+d(u)\bigr]\partial_z
+ \bigl\{\bar z[2e(u)-a(u)-c(u)]+f(u)+tg(u)e^{-u}\bigr\}
\partial_{\bar z}
\nonumber
\end{eqnarray}
 Since the solution itself
depends only on two arbitrary functions $\alpha(u),h(u)$, it is
clear that in the set (\ref{symsol}) there exist such symmetries
with respect to which the solution (\ref{spainsol}) is invariant
for any fixed $\alpha(u),h(u)$. Those symmetries are not
symmetries of the Boyer-Finley equation, therefore
(\ref{spainsol}) determines conditionally invariant (though
non-invariant) solution by the definition of Levi and Winternitz
\cite{levwin}.

In a similar way we study symmetries of the solutions
(\ref{solplus}) and (\ref{solminus}) of the Boyer-Finley equation
(\ref{heav}) obtained by Calderbank and Tod \cite{tod} and in our
paper \cite{msw}. They are general solutions of the system of PDEs
consisting of the original equation (\ref{heav}) and the
differential equations
\[ u_{z\bar zt} = 0,\qquad
u_{zz\bar z}=u_zu_{z\bar z}+e^uu_{zt}\left[\kappa u_t
+i\sqrt{2\kappa e^{-u}u_{z\bar z}-u_t^2}\right]\] plus complex
conjugate to the last equation, that follow from (\ref{solresol})
by using the definitions (\ref{sigtau}), (\ref{difinv}) of
$\tau,\lambda,\bar \lambda,\rho$. The symmetries of solutions
(\ref{solplus}) and (\ref{solminus}) with arbitrary (not fixed!)
functions $b(z),c(z)$ coincide with the symmetries of this system
of four PDEs which turn out to be exactly the same as symmetries
(\ref{symgen}) of the original equation (\ref{heav}). Hence, our
solutions with arbitrary fixed generic $b(z),c(z)$ have no
symmetries \cite{msw}. Therefore they are not conditionally
invariant and thus the method of group foliation seems to be the
only regular method which could give such solutions.

To prevent misunderstanding, we warn against mixing up the
invariance of a particular solution with a fixed choice of
arbitrary functions with the invariance of its orbit when
arbitrary functions are allowed to transform under a symmetry
transformation since the orbit is always an invariant manifold.

\section{Transformations of differential invariants \protect\\ and
invariant differential operators}
\setcounter{equation}{0}

Our main problem is to understand how we could arrive at the
solution (\ref{spainsol}) by the method of group foliation, {\it
i.e.} to determine the orbit of this solution.

The first step is to make the hodograph transformation of the
whole group foliation structure. In this section we make the
hodograph transformation of differential invariants and of
invariant differential operators and in the next section we
present the hodograph transformation of the resolving equations.

Since the hodograph transformation is a point transformation, it
conserves point symmetries, differential invariants and operators
of invariant differentiation. We change the definitions of
differential invariants $\rho,\eta$ and invariant differential
operators after the hodograph transformation and therefore from
now on we label the old invariants and operators with the tilde.

The set of hodograph-transformed differential invariants of up to
the second order inclusive consists of the new unknown
$t=t(u,z,\bar z)$ and its derivatives $t_u=1/u_t$,
$t_{uu}=-u_{tt}/u_t^3$ and two more invariants
\[ \rho=e^{-u}\bigl[t_zt_{\bar z}t_{uu}+t_u^2t_{z\bar z} -t_u(t_{\bar
z}t_{uz}+t_zt_{u\bar z})\bigr],\quad \tilde\rho=e^{-u}u_{z\bar
z}=-\rho/t_u^3 \]
\[ \eta=e^{-u}(t_zt_{uu}-t_ut_{uz})(t_{\bar
z}t_{uu}-t_ut_{u\bar z}),\qquad \tilde\eta=e^{-u}u_{zt}u_{\bar
zt}=\eta/t_u^6\]
The explicit invariant form of the transformed
Boyer-Finley equation (\ref{hodheav}) becomes
\begin{equation}
\label{invhodheav}
t_{uu}=\kappa\rho + t_u
\end{equation}
Let us denote by $D_u$ the total derivative with respect to $u$
taken for the constant values of $z,\bar z$. Since the invariant
differential operator $\delta=D_t=(1/t_u)D_u$ is a total
derivative with respect to invariant $t$, its definition is not
changed: $\tilde\delta=\delta$. The definitions of
$\Delta,\bar\Delta$ become ($\tilde\Delta=-(1/t_u^4)\Delta,\;
\bar{\tilde\Delta}=-(1/t_u^4)\bar\Delta$)
\[ \Delta=e^{-u}(t_{\bar
z}t_{uu}-t_ut_{u\bar z})(t_zD_u-t_uD_z),\quad
\bar\Delta=e^{-u}(t_zt_{uu}-t_ut_{uz})(t_{\bar z}D_u-t_uD_{\bar
z})\]

We choose $t,t_u,\rho$ as the independent invariant variables in
accordance with our previous choice. The action of new operators
of invariant differentiation on the independent variables becomes
\begin{eqnarray}
\!\!\! &\!\!\!\! &\!\!\! \delta(t)=1,\quad
\delta(t_u)=\kappa\frac{\rho}{t_u}+1, \quad
\Delta(t)=\bar\Delta(t)=0,\quad \Delta(t_u)=\bar\Delta(t_u)=\eta
\nonumber
\\ \!\!\! &\!\!\!\! &\!\!\!
\delta(\rho)=\tau,\quad\Delta(\rho)=\sigma,\quad\bar\Delta(\rho)=\bar\sigma
\label{deltact}
\end{eqnarray}
where for $\delta(t_u)$ we have used the equation
(\ref{invhodheav}).

New invariant differential operators $Y=(1/\eta)\Delta$ and $\bar
Y=(1/\eta)\bar\Delta$ have the properties
\begin{equation}\label{actYnew}
Y(t)=0, \quad \bar Y(t)=0,\quad Y(t_u)=\bar Y(t_u)=1, \quad
Y(\rho)=\lambda, \quad \bar Y(\rho)=\bar\lambda
\end{equation}

\section{Hodograph transformation of resolving \protect\\ equations}
\setcounter{equation}{0}

We consider the automorphic equation
\begin{equation}
\label{traut}
\eta=\Delta(t_u)=F(t,t_u,\rho)
\end{equation}
together with the transformed Boyer-Finley equation
(\ref{invhodheav}) and derive the resolving equations as
compatibility conditions of these two equations. The resulting
five equations coincide with the hodograph transform of the old
resolving equations with the only change caused by the difference
in definitions of differential invariants and invariant
differential operators
\begin{eqnarray}
&
&\delta(F)=\frac{1}{t_u}\,\bigl[\kappa(\lambda+\bar\lambda)+1\bigr]F
\nonumber
\\
& & F\bigl(Y(\bar\lambda)-\bar
Y(\lambda)\bigr)=(\lambda-\bar\lambda)\left(t_u\tau
-3\kappa\,\frac{\rho^2}{t_u}+\frac{2}{t_u}\,F-2\rho\right)\nonumber
\\ & & \delta(\lambda)=Y(\tau)
+ \kappa\,\frac{\lambda}{t_u}\left(\frac{\rho}{t_u}-\lambda\right)
\label{hodres}
\\[1mm] & &\delta(\bar\lambda)=\bar Y(\tau) +
\kappa\,\frac{\bar\lambda}{t_u}\left(\frac{\rho}{t_u}-\bar\lambda\right)
\nonumber
\\[1mm]
& & F\bigl(Y(\bar\lambda)+\bar Y(\lambda)\bigr)=(\lambda+\bar\lambda)
\left(-t_u\tau+4\,\frac{F}{t_u}+3\kappa\,\frac{\rho^2}{t_u}+2\rho\right)
\nonumber
\\[1mm] & &\mbox{}+2\kappa t_u^2\delta(\tau)-(12\rho+4\kappa t_u)\left(\tau
+\frac{F}{t_u^2}-\kappa\,\frac{\rho^2}{t_u^2}-\frac{\rho}{t_u}\right)\nonumber
\end{eqnarray}
Thus, we have $4$ unknowns $F,\lambda,\bar\lambda$ and $\tau$ and
$3$ independent variables $t,u_t,\rho$. Here $\delta,Y,\bar Y$
denote the projected operators (see (\ref{deltact}),
(\ref{actYnew}))
\begin{equation}
\label{deltapro}
\delta=\partial_t+\left(\kappa\,\frac{\rho}{t_u}+1\right)\partial_{t_u}
+\tau\partial_\rho,\quad Y=\partial_{t_u}+\lambda\partial_\rho,\quad
\bar Y=\partial_{t_u}+\bar\lambda\partial_\rho
\end{equation}

\section{Towards the invariant description of the hodograph solution}
\setcounter{equation}{0}

Our final goal is to obtain a complete set of independent
additional relations between differential invariants which are
satisfied on the hodograph solution of \cite{manal}. Combining
them with the resolving equations we would arrive at the
particular solution of the resolving system which selects the
orbit of this solution and makes it possible to reconstruct the
solution (\ref{spainsol}). Using these invariant relations with
the opposite sign of $\kappa$, we hope to obtain new non-invariant
solutions of the Euclidean Boyer-Finley equation (\ref{heav}) with
$\kappa=-1$ which are more physically interesting.

Up to now we have succeeded to discover only three additional
relations between differential invariants satisfied identically on
the solution (\ref{spainsol})
\begin{equation}
\label{eq1}
t_u\tau+\frac{\Phi}{t_u}=\frac{1}{4}\,\bigl[(2\rho+t_u)(\lambda+\bar\lambda)
-i\sqrt{4\Phi-t_u^2}(\lambda-\bar\lambda)\bigr],
\end{equation}
\begin{eqnarray}
\Delta(\Phi)& = &\frac{i}{4}\,\sqrt{4\Phi-t_u^2}\left\{\frac{}{} 2
\Bigl[\Phi(\lambda+\bar\lambda)+(\rho^2+t_u\rho)(\lambda-\bar\lambda)\Bigr]\right.\nonumber
\\ & &\left. -\Bigl[t_u^2 + i(2\rho+t_u)\sqrt{4\Phi-t_u^2}\,\,\Bigr]\bar\lambda\right\}
+ \frac{2}{t_u}\,\Phi F
\label{eq2}
\end{eqnarray}
together with the complex conjugate to (\ref{eq2}) where
$\Phi=F-\rho^2-t_u\rho$. We use the first one of these equations
to exclude $\tau$ from the remaining equations. It turns out that
these additional equations are not sufficient for solving the
resolving equations (\ref{hodres}). We are now in the process of
searching for missing relations.

\section{Conclusions and outlook}

In the first part of this paper we have given a short exposition
of our development of the method of group foliation as a method
for obtaining non-invariant solutions of PDEs which admit an
infinite dimensional symmetry group. The method was illustrated by
the physically interesting Boyer-Finley (or heavenly) equation. It
was shown how the constraint of commutativity of two operators of
invariant differentiation led to non-invariant solutions of the
Boyer-Finley equation.

In the second part of the paper we discussed the hodograph
solution of Ma\~nas and Alonso. We have shown how it naturally
arises after making the hodograph transformation of the
Boyer-Finley equation without any reference to the
hydrodynamic-type systems. It turns out that this solution has
extra symmetries as compared to the equation itself and that, as a
consequence, it is conditionally invariant, unlike non-invariant
solutions obtained previously.

Our final goal in this part of the paper was to obtain a complete
set of independent additional relations between differential
invariants which are valid for the hodograph solution of
\cite{manal} and using them to obtain a particular solution of the
resolving system which selects the orbit of the hodograph
solution. In this way we hoped to obtain more general solutions of
resolving equations if we would skip one of the additional
relations between differential invariants and, as a consequence,
more general solutions of the Boyer-Finley equation in a hodograph
form. More than that, those extra relations should not be
sensitive to the sign of $\kappa$ as it was in the case of
solutions (\ref{solplus}) and (\ref{solminus}) and hence we could
reconstruct solutions of the equation (\ref{hodheav}) with the
opposite sign $\kappa=-1$ which was impossible using the method of
the paper \cite{manal}. Until now we have not yet succeeded to
fulfil this program and present here only partial results.

\vspace{2mm}
{\bf Acknowledgements.} I thank Evgenii Ferapontov
and Yavuz Nutku for useful discussions.

\end{document}